\documentclass[12pt,paper]{JHEP}
\usepackage{epsfig}
\usepackage{amssymb}
\let\a=\alpha   \let\b=\beta   \let\g=\gamma   \let\d=\delta
\let\e=\epsilon    \let\h=\eta     
      \let\l=\lambda  \let\m=\mu
\let\n=\nu           \let\p=\pi      
\let\s=\sigma  \let\t=\tau      \let\f=\phi
\let\c=\chi     \let\y=\psi    
\let\G=\Gamma     
           
           \let\W=\Omega

\newcommand{\Realint}{\mathbb R}
\newcommand{\be}{\begin{equation}}
\newcommand{\ee}{\end{equation}}
\newcommand{\bea}{\begin{eqnarray}}
\newcommand{\eea}{\end{eqnarray}}
\newcommand{\ba}{\begin{array}}
\newcommand{\ea}{\end{array}}

\newcommand{\drawsquare}[2]{\hbox{%
\rule{#2pt}{#1pt}\hskip-#2pt
\rule{#1pt}{#2pt}\hskip-#1pt
\rule[#1pt]{#1pt}{#2pt}}\rule[#1pt]{#2pt}{#2pt}\hskip-#2pt
\rule{#2pt}{#1pt}}
\newcommand{\fund}{\raisebox{-.5pt}{\drawsquare{6.5}{0.4}}}
\newcommand{\Yasymm}{\raisebox{-3.5pt}{\drawsquare{6.5}{0.4}}\hskip-6.9pt%
        \raisebox{3pt}{\drawsquare{6.5}{0.4}}}
\newcommand{\antifund}{\overline{\fund}}
\newcommand{\bYasymm}{\overline{\Yasymm}}
\usepackage{graphics}
\usepackage{amssymb} 
\usepackage{amsfonts}
\usepackage{feynmf}
\title{4D Anomalous U(1)'s, their masses and their relation to
       6D anomalies}
\author{P. Anastasopoulos
\\
Department of Physics, University of Crete\\
P.O.Box 2208, GR-710 03 Heraklion, GREECE\\
\smallskip
and \\
\smallskip
Laboratoire de Physique Th{\'e}orique Ecole
Polytechnique \\
91128, Palaiseau, FRANCE\\
Email: {\tt panasta@physics.uoc.gr}}
\preprint{\hepth{0306042}\\ CPHT-RR 026.0503}

\abstract{In some four-dimensional orientifolds, $U(1)$ gauge
fields that are free of four-dimensional anomalies can still be
massive. It is shown that this is due to mass-generating
six-dimensional anomalies. Six-dimensional anomalies affect
four-dimensional masses via decompactifications.}

\begin{document}

\maketitle

\section{Introduction}

Recently, many attempts have been made in order to embed the
Standard Model (SM) in open string theory, with some success
\cite{Aldazabal:2000dg, Ibanez:2001nd, Blumenhagen:2000ea,
Cvetic:2002qa, Bailin:2000kd, Kokorelis:2002ip,
Antoniadis:2000en}. They consider the SM particles as open string
states attached on different stacks of D-branes. $N$ coincident
D-branes typically generate a Unitary group $U(N)$. Therefore,
every stack of branes supplies the model with extra abelian gauge
fields.

Such $U(1)$ fields have generically four-dimensional anomalies.
Such anomalies are cancelled via the Green-Schwarz mechanism
\cite{Green:sg, Sagnotti:1992qw, Ibanez:1998qp} where a scalar
axionic field (zero-form, or its dual two-form) is responsible for
the anomaly cancellation. This mechanism gives a mass to the
anomalous $U(1)$ fields and breaks the associated gauge symmetry.

If the string scale is around a few TeV, observation of such
anomalous $U(1)$ gauge bosons becomes a realistic possibility
\cite{Kiritsis:2002aj, Ghilencea:2002da, Ghilencea:2002by}.

As it has been shown in \cite{Antoniadis:2002cs}, we can compute
the masses of the anomalous $U(1)$s by evaluating the ultraviolet
tadpole of the one-loop open string diagram with the insertion of
two gauge bosons on different boundaries. In this limit, the
diagrams of the annulus with both gauge bosons in the same
boundary and the M\"obius strip do not contribute when vacua have
cancelled tadpoles.

It turns out that $U(1)$ gauge fields that are free of
four-dimensional anomalies can still be massive
\cite{Ibanez:2001nd, Scrucca:2002is, Antoniadis:2002cs}. We will
show that this is due to the presence of mass-generating
six-dimensional anomalies. Since there are decompactification
limits in the theory, six-dimensional anomalies affect
four-dimensional masses.

In six dimensions, two type of fields are necessary to cancel the
anomalies, a scalar axion and a two-form. There is also a
four-form field but it is dual to the scalar. Via the
Green-Schwarz mechanism, the pseudoscalar axions give mass to the
anomalous $U(1)$ fields. However, the two-forms are not involved
in mass generation.


In this paper, we show that four-dimensional non-anomalous $U(1)$s
can have masses if their decompactification limits suffer by
six-dimensional anomalies. We calculate the masses of the
anomalous $U(1)$s of various six-dimensional orientifolds and we
compare our results with decompactification limits of the
four-dimensional orientifolds $Z'_6$ and $Z_6$.

The paper is organized as follows. In Section 2, we describe the
structure of the six-dimensional mixed gauge anomalies.
In Section 3, we present the one-loop string computation for the
mass formula of the anomalous $U(1)$ in six dimensions. After, we
give some examples of N=1 six-dimensional orientifolds where we
provide the effective field theory predictions about the
anomalous $U(1)$s and we evaluate the mass of these anomalous
$U(1)$s using the formulas that we found before.
In Section 4, various decompactification limits of
four-dimensional orientifold $Z'_6$ and $Z_6$ are studied and
compared with six-dimensional orientifolds.

\section{The structure of six-dimensional mixed gauge anomalies}

In six dimensions, the leading diagram that can give a
contribution to anomalies is the square diagram (it has 1+D/2
external gauge bosons) \cite{Sagnotti:1992qw}.
In the presence of an anomalous $U(1)$ field, the effective action
is not invariant under a transformation $\d A^i = d \e^i$. In six
dimensions, the only possible non-zero mixed-anomaly diagrams are:
\be \d S|_{gauge} = \int d^6x \left[ Tr[Q_i Q_j T^\a T^\a]
~\e^i~F^j\wedge Tr[G^2] + Tr[Q_i T^\a \{T^\b T^\g\}]~\e^i~ Tr[G^3]
\right] \label{dS}\ee
where powers of forms are understood as wedge products. We denote
by $G$ the field strength of a non-abelian gauge field $W_\m$.
Gauge invariance is preserved by some other terms in the effective
action that cancel the anomalous variations. The cancellation of
the first anomalous term is arranged by a 2-form $B^i$ which
transform under the $U(1)$ transformation like $\d B^i=-\e^i F^i$:
\be S_{QQTT} =\int d^6x \left[-{1 \over 4 g_i^2} F^{i2}_{\m\n}
-{1\over 12} \left[d B^i + \W_{A^i} \right]^2 + C_1~B^i \wedge
Tr[G^2] \right] \label{S_QQTT}\ee
where $C_1=Tr[Q_i Q_j T^\a T^\a]$ is the anomaly of the first
diagram and the 3-form $\W_{A}=AdA$ is the Chern-Simons term of
the abelian gauge field $A^i_\m$. This part of the action does not
generate a mass for the gauge boson.

By the (\ref{S_QQTT}), we can evaluate the action in terms of the
dual 2-form $\l$ of $B$ \cite{Ghilencea:2002da}. Using $Tr[G_i
\tilde{G}_i]=d \W_{W_i}$, where $\W_{W}=Tr[WdW+{2\over 3}W^3]$ is
the Chern-Simons term for the non-abelian gauge field $W^i$, we
finally find:
\be \tilde{S}_{QQTT} =\int d^6x \left[-{1 \over 4 g_i^2}
F^{i2}_{\m\n} - {1\over 12}\left[d \l^i - 6 C_1~\W_{W^i} \right]^2
- {1\over 6}\W_{A^i}\wedge ( d\l^i - 6C_1~\W_{W^i}) \right]~.
\label{dual_S_QQTT}\ee
The $\l^i$ are invariant under $U(1)$ gauge transformations and
transform like $\d\l^i=6C_1~Tr[G \e^i]$ under a non-abelian gauge
transformation $\d W_\m=D_\m \e$ so that the action is gauge
invariant.

Thus, under a $U(1)$ gauge transformation the variation of
$\W_{A^i}\wedge d\l^i$ (since $\d\W_{A^i}=d\e F$) vanishes due to
integration by parts and the term $C_1~\W_{A^i} \wedge \W_{W^i}$
cancels the first anomaly in (\ref{dS}).

The second anomaly is cancelled by a pseudoscalar axion that
transforms under the $U(1)$ transformation as $\d\a^i=-\e^i$:
\be S_{QTTT} =\int d^6x \left[-{1 \over 4 g_i^2} F^{i2}_{\m\n} +
{M^2\over 2} (A^i +d \a^i)^2 + C_2~\a^i~ Tr[G^3]\right]
\label{S_QTTT}\ee
where $C_2=Tr[Q_i T^\a \{T^\b T^\g\}]$ is the anomaly of the
second diagram. This action supplies a mass term for the $U(1)$
gauge field and breaks the gauge symmetry in six dimensions.
\vspace{.6cm}
\FIGURE[ht]{
\unitlength=0.6mm
\begin{fmffile}{Q2G2}
\begin{fmfgraph*}(40,30)
\fmfpen{thick} \fmfleft{i1} \fmfright{o1}
\fmftop{v1,v2,v7,v8,v9} \fmfbottom{v4,v5,v10,v11,v12}
\fmf{plain}{i1,v2,v3,v5,i1}
\fmf{gluon}{v3,o1}
\fmffreeze \fmfdraw
\fmf{photon}{v2,v8}
\fmf{gluon}{v5,v11}
\fmfv{decor.shape=circle, decor.filled=empty, decor.size=.20w}{i1}
\fmffreeze \fmfdraw \fmfv{d.sh=cross,d.size=.20w}{i1}
\fmffreeze
\fmflabel{$U(1)_i~$}{i1} \fmflabel{$U(1)_j$}{v8}
\fmflabel{$G^\a$}{v11} \fmflabel{$G^\a$}{o1}
\end{fmfgraph*}
\end{fmffile}
~~~~~~~~~~~~~~~~~~~~~~
\unitlength=0.6mm
\begin{fmffile}{QG3}
\begin{fmfgraph*}(40,30)
\fmfpen{thick} \fmfleft{i1} \fmfright{o1}
\fmftop{v1,v2,v7,v8,v9} \fmfbottom{v4,v5,v10,v11,v12}
\fmf{plain}{i1,v2,v3,v5,i1}
\fmf{gluon}{v3,o1}
\fmffreeze \fmfdraw
\fmf{gluon}{v2,v8}
\fmf{gluon}{v5,v11}
\fmfv{decor.shape=circle, decor.filled=empty, decor.size=.20w}{i1}
\fmffreeze \fmfdraw \fmfv{d.sh=cross,d.size=.20w}{i1}
\fmffreeze
\fmflabel{$U(1)_i~$}{i1} \fmflabel{$G^\a$}{v8}
\fmflabel{$G^\g$}{v11} \fmflabel{$G^\b$}{o1}
\end{fmfgraph*}
\end{fmffile}
\vspace{.6cm}
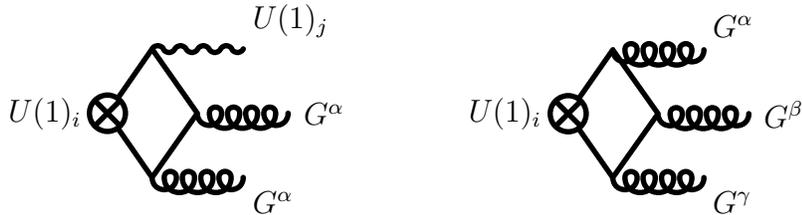
\caption{The anomalous diagrams are squares in six dimensions. The
only mixed-gauge diagrams that are anomalous are $Tr[Q_i Q_j T^\a
T^\a]$ and $Tr[Q_i T^\a T^\b T^\g]$.}}

\section{Calculation of the mass of the anomalous $U(1)$s
         for six-dimensional orientifolds}

In this section we will evaluate the contribution to the anomalous
$U(1)$ mass for six-dimensional supersymmetric orientifolds. These
models appear as decompactification limits of four-dimensional
orientifolds.

In Type I string theory, the axions that are relevant for anomaly
cancellation come from the RR sectors. The mass-term in
(\ref{S_QTTT}) is coming from different orders of string
perturbation theory. The $(\partial \a^i)^2$ is a tree-level
(sphere) term, the $A^i\partial \a^i$ comes in the disk and the
quadratic term in the gauge fields is a one-loop contribution. To
clarify this, we mention that $g_i^2$ is proportional to
$g_s=e^\f$ and every power of the axion absorbs a dilaton factor
$e^{-\f}$ because it is a RR filed. The string perturbation series
are weighted by $g_s^{-\c}$ where $\c=2-2h-c-b$ is the Euler
character and $h$, $c$ and $b$ denote the handle, the cross-cups
and the boundaries of a closed orientable Riemann surface
respectively.

The diagrams at one-loop that contribute to terms quadratic in the
gauge bosons (anomalous $U(1)$s) are the genus-one surfaces with
boundaries: the annulus and the M\"obius strip.
In the infrared (IR) region they diverge logarithmically and give
the logarithmic running of the couplings. In the ultraviolet (UV)
region the tadpoles of the annulus with both gauge bosons inserted
in the same boundary and the M\"obius strip vanish due to the
tadpole cancellation. However, in this UV limit the annulus
amplitude with the gauge bosons inserted in opposite boundaries
provides the mass-term of the anomalous
$U(1)$\cite{Antoniadis:2002cs}. Since we are interested in the
anomalous gauge boson mass, we concentrate on the latter diagram.
The gauge boson vertex operator is
\be \tilde{V}^a=\l^a \e_\m (\partial X^\m +i(p\cdot \y)\y^\m)
e^{ip\cdot X} \ee
where $\l$ is the Chan-Paton matrix and $\e^\m$ is the
polarization vector. The 2-point annulus amplitude is given by
\be {\cal A}^{ab}=-{1\over 4G}\int [d\t][dz] \int {d^6p \over
(2\p)^6}\sum_k \langle \tilde{V}^a(\e_1,p_1,z) \tilde{V}
^b(\e_2,p_2,z_0) \rangle_k \label{Annulus11} \ee
where $G$ denotes the order of the orientifold group. The
fundamental polygon of the annulus is $[0,t/2]\otimes[0,1/2]$. The
index $k$ denotes the various orbifold sectors that we may have.
Using the translation symmetry of the annulus, we fix the position
of one VO to $z_0=1/2$. The other VO is placed on the imaginary
axis with $z\in [0,t/2]$.

The leading term of (\ref{Annulus11}) is
\be {\cal A}^{ab}=\int {d^6p \over (2\p)^6} [(\e_1\cdot
\e_2)(p_1\cdot p_2)-(\e_1\cdot p_2)(p_1\cdot \e_2)] \sum_k Tr[\g_k
\l^\a] Tr[\g_k \l^\b] {\cal A}^{ab}_k. \label{Annulus11-2} \ee
where
\be {\cal A}^{ab}_k=-{1\over 2G}\int [d\t][dz] e^{-p_1\cdot p_2
\langle X(z) X(1/2)\rangle} \left[ \langle \y(z)\y(1/2)\rangle^2-
\langle X(z)\partial X(1/2)\rangle^2 \right]Z^{ab}_k.
\label{Annulus11ab} \ee
since the $p$-independent terms vanish due to supersymmetry. The
bosonic and fermionic correlation functions are given in the
Appendix (\ref{PartialBosonicAC}), (\ref{FIdentityTapp}).

It appears that the amplitude (\ref{Annulus11}) has a kinimatical
multiplicative factor that is ${\cal O}(p^2)$, thus would seem to
provide a leading  correction only to the anomalous gauge boson
coupling. We will see however, that after integration over the
position $z$ and the annulus modulus $\t_2$, a term proportional
to $1/ p_1\cdot p_2$ appears from the ultraviolet (UV) region (as
a result of the quadratic UV divergence in the presence of
anomalous $U(1)$s) that will provide the mass-term.

Strictly speaking, the amplitude above is zero on-shell if we
enforce the physical state conditions $\epsilon\cdot p=p^2=0$ and
momentum conservation $p_1+p_2=0$. There is however a consistent
off-shell extension, without imposing momentum conservation, that
has given consistent results in other cases (see
\cite{Kiritsis:1997hj} for a discussion) and we adopt it here. We
will thus impose momentum conservation only at the end of the
calculation.

Spin structure summation of the partition function $Z^{ab}_k$,
gives zero due to space-time supersymmetry. Therefore, terms in
the correlation functions which are spin-structure independent
vanish. The only spin-dependant term lies in the fermionic
correlation function:
\be \langle \y(z-1/2)\y(0)\rangle^2[^\a_\b]_{annulus} = -2\p
i\partial_\t \log \vartheta[^\a_\b](0|\t). \label{y(z)y(o)} \ee
Equ. (\ref{y(z)y(o)}) is independent of $z$, the position of the
second VO. Thus, we can easily integrate on $dz$. Using the modular
transformations of the theta functions and keeping the leading
order of $\d$, we have:
\bea \int_0^{\t_2} dz ~e^{-\d \langle X(z)X(0)\rangle}=
\int_0^{\t_2} dz ~\t_2^{\d/2} {(2\p\h^3(\t))^\d \over
\vartheta[^0_1](z/\t|-1/\t)} =\t_2^{1+\d/2}[2\p \h^3(\t)]^\d+...
\eea
Following the procedure of \cite{Antoniadis:2002cs} we rewrite
(\ref{Annulus11ab}) as:
\be {\cal A}^{ab}_k=-{1\over 2G}\int [d\t] \t_2^{1+\d/2}[2\p
\h^3(\t)]^\d F^{ab}_k. \label{Annulus11ab-1} \ee
defining $F^{ab}_k$ as a term which contains all the
spin-structure and the orbifold information:
\be F^{ab}_k= \sum_{\a\b} \h^{\a\b} \left[-2\p i \partial
\log\vartheta[^\a_\b]\right] \left[ {1\over (2 \p
\t)^3}{\vartheta[^\a_\b]^2\over \h^6} \right]
Z_{int,k}^{ab}[^\a_\b] \label{Fab-k}\ee
where $\h^{\a\b}={1\over 2}(-1)^{\a+\b+\a\b}$. The first bracket
is denoting the VO insertion in the annulus diagram. The second is
the six-dimensional partition function.
\TABLE[h]{\footnotesize
\renewcommand{\arraystretch}{1.25}
\begin{tabular}{|c|c|c|}
\hline
Twist Group   & & \\
\cline{1-1} Gauge Group & \raisebox{2.5ex}[0cm][0cm]{ (99)/(55)
matter} &
\raisebox{2.5ex}[0cm][0cm]{ (95) matter}  \\
\hline\hline $Z_2 $ & $2 \times 120+ 2 \times \overline{120} $  &
$(16; \overline{16})+(\overline{16}; 16)$  \\
\cline{1-1}
$U(16)_9\times U(16)_5$ &  & \\
\hline\hline $Z_3$ &$(8,16_v)+
(\overline{8},16_v)$ &  -  \\
\cline{1-1}
$U(8)\times SO(16)$ & $ +(28,1)+ (\overline{28},1) $   & \\
\hline\hline $Z_4 $ & $(28,1)+ (\overline{28},1) $  &
$(8,1;\overline{8},1)+(\overline{8},1;8,1)$  \\
\cline{1-1} $U(8)_9\times U(8)_9\times$ & $ +(1,28)+
(1,\overline{28}) $ &
$ +(1,8;1,\overline{8})+(1,\overline{8};1,8) $\\
$U(8)_5\times U(8)_5$ & $ +(8,\overline{8}) + (\overline{8},8) $ & \\
\hline\hline $Z_6$ &$(\underline{6, 1}, 1)
+(\underline{\overline{6},1},1)$
 &  $(4,1,1;\overline{4},1,1)+(\overline{4},1,1;4,1,1)$ \\
\cline{1-1} $( U(4)^2\times U(8))_9\times$ & $+
(\underline{4,1},\overline{8})+(\underline{\overline{4},1},8)$
 & $+(1,4,1;1,\overline{4},1)+(1,\overline{4},1;1,4,1)$  \\
$+( U(4)^2\times U(8))_5$    &
& $+(1,1,8;1,1,\overline{8})+(1,1,\overline{8};1,1,8)$ \\
\hline
\end{tabular}
\caption{The transformations of the massless fermionic states in
all the D=6 orientifolds. The underlined numbers denote all the
possible permutations.}}

\subsection{Six-dimensional N=1 orientifolds examples}

Usual six-dimensional decompactification limits of
four-dimensional supersymmetric orientifolds are the N=1
orientifolds of Type IIB string theory, $\Realint^6\times K3/Z_N$ where
the only possible choices are $N=2,3,4,6$. Thus, we will apply the
above general formulae on these orientifolds.

We re-evaluate the massless spectrum of these models using the
'shift' vectors that are given for each model. The result is
provided in Table 1. We were especially careful in distinguishing
the representations from the conjugate representations since this
was not transparent in the previous literature
\cite{Bianchi:1990tb, Gimon:1996rq, Gimon:1996ay}. Tadpole
cancellation guaranties that the models are free of irreducible
non-Abelian anomalies \cite{Aldazabal:1999nu, Sagnotti:1992qw}.
This is also shown in appendix \ref{AnomalyFree}.

The mixed-anomaly traces can be easily evaluated for each
orientifold. Our normalization of the cubic casimir and the $U(1)$
charge of the $SU(N)$ representations are given in Table 2.
\TABLE[h]{
\begin{tabular}{ccc}
$SU(N)$ Representation & Cubic Casimir & $U(1)$ Charge \\
\hline
$\fund $ &  $A(\fund)=1 $ & $Q(\fund)=1$ \\
$\antifund$ & $A(\antifund)=-1 $ & $Q(\antifund)=-1$ \\
$\Yasymm $ & $A(\Yasymm)=N-4 $ & $Q(\Yasymm)=2$ \\
$\bYasymm$ & $A(\bYasymm)=-N+4 $ & $Q(\bYasymm)=-2$ \\
\end{tabular}\caption{Our normalization of the cubic casimir and the $U(1)$
charge of the $SU(N)$ representations.}}

The general mass formulae for the anomalous $U(1)$ gauge fields in
the orientifolds above can be easily evaluated. More details for
the explicit computations of the UV tadpole in the annulus diagram
(\ref{Annulus11}) are provided in the Appendix
\ref{AnnulusUVchapter}. The results for strings attached on the
same kind of branes (untwisted states) are
(\ref{Annulus11ab-UV-A})
\be {1\over 2}M^2_{aa}= -{4 \over \p^2 N} \sum_k \sin^2 {\p k\over
N} ~ Tr[\g_k \l^{a}] Tr[\g_k \l^{a}] \label{Mass_aa}\ee
where $a=5,9$ denotes the kind of D-branes on which the open
string is attached. In the case where strings have one end on a
$D5$ and the other on a $D9$-brane (twisted states) we
have\footnote{We could end up in the same results (\ref{Mass_aa},
\ref{Mass_59}) if we made use of the axionic couplings for the
six-dimensional orientifolds, evaluated first for the $Z_2$ case
in \cite{Berkooz:1996iz} and after for all $Z_N$ in
\cite{Scrucca:1999pp}.}:
\be {1\over 2}M^2_{59}= -{1 \over \p^2 N} \sum_k Tr[\g_k \l^{5}]
Tr[\g_k \l^{9}]~. \label{Mass_59}\ee

We should mention, that the above masses are unormalized. To
obtain the normalized mass matrix, we must also take into account
the kinetic terms of the $U(1)$ gauge bosons which are
\be S_{\rm kinetic}=-{1\over 4g_s}\left[{\cal V}_1{\cal V}_2
\sum_i F_i^2 +\sum_j
\tilde F_j^2\right]\, . \ee
where $i$ and $j$ denote the gauge groups that are coming from
different stacks of D9 and D5-branes. This implies $M^2_{99}\to
M^2_{99}/({\cal V}_1{\cal V}_2)$, $M^2_{55}\to M^2_{55}$ and $M^2_{95}\to
M^2_{95}/(\sqrt{{\cal V}_1{\cal V}_2})$.

A convenient formula of the action of the orbifold elements on the
Chan-Paton factors is defined by the matrices:
\be \g_k= e^{-2\p i k V\cdot H} \label{g_on_CP}\ee
where $H_I$ the Cartan generators of $SO(32)$, represented by
diagonal matrices having the $\s^3$ Pauli matrix in the diagonal
and everywhere else zero. Notice that the normalization of the
Cartans is $Tr[H_I^2]=2$. $V^I$ is a 16-dimensional "shift"
vector, defined by (\ref{g_on_CP}) and satisfies the tadpole
conditions for each orientifold.

The normalized generators of the anomalous $U(1)_i$ are defined
as:
\be \l^\a_i = {1\over 2\sqrt{n_i}}\sum Q^\a_i \cdot H
\label{lambda_Gen}\ee
where $\a$ denotes the type of brane. The $Q^\a_i=
(0,\ldots,0,1,\ldots,1,0,\ldots,0)$ is a 16-dimensional vector
with $n_i$ entries of 1s where the $SU(n_i-1)$ lives. We normalize
the $\l$ matrices with $Tr[\l^2]=1/2$. Thus, the relevant trace
is:
\be Tr[\g^\a_k \l^\a_i]=Tr[e^{-2\p i k V^\a \cdot H} Q^\a_i \cdot
H] =- {i\over \sqrt{n_i}} \sin[2\p k V^\a_i] \label{Trace[gl]}\ee
where $V^\a_i$ are the overlapping components of $V^\a$ and $Q^\a$
\cite{Ibanez:1998qp}.

\subsubsection{$Z_2$ orientifold}\label{Z_2}

For the $Z_2$, the tadpole condition gives 32 $D9$ and 32
$D5$-branes \cite{Bianchi:1990tb, Gimon:1996rq, Gimon:1996ay}. The
characteristic vectors are:
\be V_{5,9}={1\over 4}(1,1,1,1,1,1,1,1,1,1,1,1,1,1,1,1)~. \ee
The gauge group is $U(16)_9 \times U(16)_5$. The massless states
are given in Table 1 and we use them to evaluate the mixed
anomalous diagrams. We are interested in anomalous diagrams with
one abelian and three non-abelian gauge bosons $U(1)\times
SU(N)^3$ since their cancellation provides the six-dimensional
mass-term. We find:
\be A_{QTTT} =
32 \cdot \left( \ba {cc}
 4 & -1 \\
-1 &  4 \ea \right). \ee
where the columns label the U(1)s, while the rows label the
non-abelian factors. The matrix has two non-zero eigenvalues and
both anomalous $U(1)$s are expected to become massive
\cite{Scrucca:1999pp}. The unormalized mass matrix for the
anomalous $U(1)$s is calculated by the use of (\ref{Mass_aa}),
(\ref{Mass_59}) and (\ref{Trace[gl]}):
\be {1\over 2}M^2=
-{1 \over 2\p^2}\left( \ba {cc}
4~Tr[\g_1 \l^{9}] Tr[\g_1 \l^{9}] & Tr[\g_1 \l^{9}] Tr[\g_1\l^{5}] \\
Tr[\g_1 \l^{5}] Tr[\g_1 \l^{9}] & 4~Tr[\g_1 \l^{5}] Tr[\g_1\l^{5}]
\\ \ea \right)
={8\over \p^2}\left( \ba {cc}
4 & 1\\
1 & 4\\\ea \right). \label{Z_2massOfU(1)}\ee
As it was expected from the effective field theory computation of
the anomalies, there are two massive eigenstates: $\pm A
+\tilde{A}$ with masses $24/\p^2$, $40/\p^2$ (we denote with $A$
the gauge boson that is coming from the $D9$-branes and with
$\tilde{A}$ the one that is coming from the $D5$).

\subsubsection{$Z_3$ orientifold}\label{Z_3}

The $Z_3$ orientifold does not contain a $Z_2$ reflection element.
Thus, there are no $D5$-branes. The characteristic vector is:
\be V_9={1\over 3}(1,1,1,1,1,1,1,1,0,0,0,0,0,0,0,0) \ee
and the gauge group $U(8)\times SO(16)$. From the massless
spectrum which is provided in Table 1 we find that the single
gauge boson suffers from mixed non-abelian anomalies
\cite{Scrucca:1999pp}.
\be A_{QTTT} = 48 . \ee
Using (\ref{Trace[gl]}) we find the mass of this gauge boson:
\be {1\over 2}M^2=~ {32 \over 3\p^2} \sum_{k=1}^2 \sin^2 {\p k
\over 3} \sin^2{2\p k \over 3} =~ {12 \over \p^2}~.
\label{Z_3massOfU(1)}\ee

\subsubsection{$Z_4$ orientifold}

The $Z_4$ orientifold contains 32 $D9$-branes and 32 $D5$-branes.
The characteristic vectors are:
\be V_{5,9}={1\over 8}(1,1,1,1,1,1,1,1,3,3,3,3,3,3,3,3) \ee
and the gauge group is $U(8)_9\times U(8)_9\times U(8)_5\times
U(8)_5$. The $U(1)\times SU(N)^3$ anomalies are:
\be A_{QTTT} =
16 \cdot\left( \ba {cccc}
 3 & -1 & -1 &  0   \\
-1 &  3 &  0 & -1   \\
-1 &  0 &  3 & -1   \\
 0 & -1 & -1 &  3   \ea \right). \ee
where again the columns label the U(1)s and the rows the
non-abelian factors $SU(8)_9^2\times SU(8)_5^2$. Notice that we
have two equal matrices in the diagonal blocks and two other ones
equal in the off-diagonal blocks. This is a consequence of the
fact that the $D9$ and $D5$ branes are related by T-duality and
split in isomorphic groups. All those models are
T-selfdual\footnote{Except from the $Z_3$ orientifold which is
T-dual to a $Z_6'$ model that does not contain the pure $\Omega$
element \cite{Gimon:1996rq}].}. The anomaly matrix has four
non-zero eigenvalues \cite{Scrucca:1999pp}.

The mass matrix of the anomalous U(1) masses is
\be {1\over 2}M^2={4\over \p^2}\left( \ba {cccc}
 3 & -1 &  1 &  0   \\
-1 &  3 &  0 &  1   \\
 1 &  0 &  3 & -1   \\
 0 &  1 & -1 &  3   \ea \right)
\ee
Diagonalizing this matrix, we find four massive $U(1)$ fields that
are in accordance with the anomalies. The massive $U(1)$ fields
are $-A_1-A_2+\tilde{A}_1+\tilde{A}_2$, $A_1+\tilde{A}_2$,
$A_2+\tilde{A}_1$, $-A_1+A_2-\tilde{A}_1+\tilde{A}_2$ with masses
$4/ \p^2$, $12/ \p^2$, $12/ \p^2$, $20/ \p^2$ respectively.

\subsubsection{$Z_6$ orientifold}

The $Z_6$ orientifold contains 32 $D9$-branes and 32 $D5$-branes.
The characteristic vectors are:
\be V_{5,9}={1\over 12}(1,1,1,1,5,5,5,5,3,3,3,3,3,3,3,3) \ee
and the gauge group $U(4)_9\times U(4)_9\times U(8)_9\times
U(4)_5\times U(4)_5\times U(8)_5$.

The $U(1)\times SU(N)^3$ anomalies are:
\be A_{QTTT} =
8 \cdot \left( \ba {cccccc}
 3 &  0 & -2 & -1 &  0 &  0 \\
 0 &  3 & -2 &  0 & -1 &  0 \\
-1 & -1 &  4 &  0 &  0 & -2 \\
-1 &  0 &  0 &  3 &  0 & -2 \\
 0 & -1 &  0 &  0 &  3 & -2 \\
 0 &  0 & -2 & -1 & -1 &  4 \ea \right).
\label{AnomalyQTTT_Z4}\ee
The columns are the U(1)s and the rows the non-abelian factors,
always in the ordered form of Table 1. The (\ref{AnomalyQTTT_Z4})
has five non-zero and one zero eigenvalue which corresponds to
$A_1+A_2+A_3+\tilde{A}_1+\tilde{A}_2+\tilde{A}_3$. Our result is
in accordance with \cite{Scrucca:1999pp} where it had been shown
that one of the six $U(1)$ factor remains unbroken. The
independent axions that participate in the cancellation of the
anomaly and the mass generation are only five.

The mass matrix for the anomalous $U(1)$s is
\be {1\over 2} M^2={2\over \p^2}\left( \ba {cccccc}
3 & 0&  -\sqrt{2} &  1 &  0&  0 \\
0 & 3&  -\sqrt{2} &  0 &  1&  0 \\
-\sqrt{2} & -\sqrt{2}& 4 & 0 & 0 & 2 \\
1 & 0&  0 & 3 & 0 & -\sqrt{2} \\
0 & 1&  0 & 0 & 3 & -\sqrt{2} \\
0 & 0&  2 & -\sqrt{2}& -\sqrt{2}& 4 \ea \right) \ee
Diagonalizing the mass matrix, we find that five $U(1)$ fields
become massive and one remains massless. The effective field
theory computation agrees with the result above.

\section{The decompactification limits of four-dimensional N=1
         orientifolds}

We are interested in studying the decompactification limits of
four-dimensional orientifolds, in order to investigate potential
six-dimensional anomalies. In this section, we focus on the $Z'_6$
and $Z_6$ four-dimensional orientifold since they have enough of
structure needed. The four-dimensional spectra of these models are
provided in Table 3.
\TABLE[h]{\footnotesize
\renewcommand{\arraystretch}{1.25}
\begin{tabular}{|c|c|c|}
\hline
Twist Group   & & \\
\cline{1-1} Gauge Group & \raisebox{2.5ex}[0cm][0cm]{ (99)/(55)
matter} &
\raisebox{2.5ex}[0cm][0cm]{ (95) matter}  \\
\hline\hline $Z'_6 $ & $(\overline{4},1,8)+
(1,4,\overline{8})+(6,1,1) $ &
$(\overline{4},1,1;\overline{4},1,1)+(1,4,1;1,4,1)$  \\
\cline{1-1} $U(4)_9^2\times U(8)_9\times$ & $ +(1,\overline{6},1)+
(4,1,8) + (1,\overline{4},\overline{8}) $ &
$ +(1,\overline{4},1;1,1,8)+(1,1,8;1,\overline{4},1) $\\
$U(4)^2_5\times U(8)_5$ & $ +(\overline{4},4,1) + (4,4,1) +
(\overline{4},\overline{4},1) $ &
$+(4,1,1;1,1,\overline{8})+(1,1,\overline{8};4,1,1)$ \\
& $ + (1,1,28) + (1,1,\overline{28}) $& \\
\hline\hline $Z_6 $ & $2 (15,1,1)+ 2(1,\overline{15},1) $ &
$(6,1,1;6,1,1)+(1,\overline{6},1;1,\overline{6},1)$  \\
\cline{1-1} $U(6)_9^2\times U(4)_9\times$ & $ +2
(\overline{6},1,4)+ 2(1,6,\overline{4}) $ &
$ +(1,6,1;1,1,\overline{4})+(1,1,\overline{4};1,6,1) $\\
$U(6)^2_5\times U(4)_5$ & $ +(\overline{6},1,\overline{4}) +
(1,6,4) + (6,\overline{6},1) $ &
$+(\overline{6},1,1;1,1,4)+(1,1,4;\overline{6},1,1)$ \\
\hline
\end{tabular}
\caption{The transformations of the massless fermionic states in
two D=4 orientifolds.}}

\subsection{The four-dimensional $Z'_6$ orientifold}

The orbifold rotation vector is $(v_1,v_2,v_3)=(1,-3,2)/6$. Since
there is an order two twist ($k=3$), we have one set of
$D5$-branes. Tadpole cancellation implies the existence of 32
$D9$-branes and 32 $D5$-branes that we put together at one of the
fixed points of the $Z_2$ action (namely the origin). The
Chan-Paton 'shift' vectors are
\be V_{5,9}={1\over 12}(1,1,1,1,5,5,5,5,3,3,3,3,3,3,3,3)~.
\label{Z6'vectors}\ee
The gauge group has a factor of $U(4)\times U(4)\times U(8)$
coming from the $D9$-branes and an isomorphic factor coming from
the $D5$-branes. Different sectors preserve different
supersymmetries. The $N=1$ sectors correspond to $k=1,5$, while
for $k=2,3,4$ we have $N=2$ sectors.

The four-dimensional anomalies of the $U(1)$s have been computed
in \cite{Ibanez:1998qp} and the anomaly matrix is
\be A_{QTT} \sim \left( \ba {cccccc}
 2 &  2 &  4\sqrt{2} & -2 &  0 & -2\sqrt{2} \\
-2 & -2 & -4\sqrt{2} &  0 &  2 &  2\sqrt{2} \\
 0 &  0 &      0     &  2 & -2 &     0 \\
-2 &  0 & -2\sqrt{2} &  2 &  2 &  4\sqrt{2} \\
 0 &  2 &  2\sqrt{2} & -2 & -2 & -4\sqrt{2} \\
 2 & -2 &      0     &  0 &  0 &     0    \ea \right) \ee
there are two linear combinations that are free of
four-dimensional anomalies: $\sqrt{2}(A_1+A_2)-A_3$ and
$\sqrt{2}(\tilde{A}_1+\tilde{A}_2)-\tilde{A}_3$.

The contribution to the mass matrix \cite{Antoniadis:2002cs} is
\bea {1\over 2}M^2_{aa,ij}&=&-{\sqrt{3}\over
24\pi^3}\left(Tr[\g_1\l^a_i]
Tr[\g_1\l^a_j]+Tr[\g_5\l^a_i]Tr[\g_5\l^a_j]\right) \nonumber\\
&-&{1\over 4\pi^3}\left( {\cal V}_2\d_{a,9}+{1\over 4{\cal
V}_2}\d_{a,5}\right)\left(Tr[\g_2\l^a_i] Tr[\g_2\l^a_j]
+Tr[\g_4\l^a_i]Tr[\g_4\l^a_j]\right)\nonumber\\
&-&{{\cal V}_3\over
3\pi^3}Tr[\g_3\l^a_i]Tr[\g_3\l^a_j]\label{Z'6.99.D4->D6}\eea
for $a=5,9$ where $\d_{a,b}$ is the Kronecker delta. The mixed 59
annulus diagrams give a contribution to the mass
\bea {1\over 2}M^2_{95,ij} =&-&{\sqrt{3}\over
48\pi^3}\left(Tr[\g_1\l^9_i] Tr[\g_1\l^5_j]+Tr[\g_5\l^9_i]
Tr[\g_5\l^5_j]\right. \nonumber\\
&+& \left. Tr[\g_2\l^9_i]Tr[\g_2\l^5_j]-
Tr[\g_4\l^9_i]Tr[\g_4\l^5_j]\right) \nonumber \\
&-& {{\cal V}_3\over 12\pi^3}Tr[\g_3\l^9_i]Tr[\g_3\l^5_j]
~.\label{Z'6.59.D4->D6}\eea

The unormalized mass matrix \cite{Antoniadis:2002cs} has
eigenvalues and eigenvectors:
\bea m^2_1=6{\cal V}_2 &~~,~~ & -A_1+A_2\, \label{MZ6'1}\\
m^2_2={3\over 2{\cal V}_2} & ~~,~~ & -\tilde A_1+\tilde A_2 \\
m^2_{3,4}={5\sqrt{3}+48{\cal V}_3\pm\sqrt{3}\a\over 12} & ~~,~~ &
{-3\pm \a\over 4\sqrt{2} (4\sqrt{3}{\cal V}_3-1)}
(A_1+A_2-\tilde A_1-\tilde A_2)-A_3+\tilde A_3;\nonumber\\
&&\\
m^2_{5,6}={15\sqrt{3}+80{\cal V}_3\pm\b \over 12} & ~~,~~ &
{9\sqrt{3}\mp\b\over 4\sqrt{2}(20{\cal
V}_3-3\sqrt{3})}(A_1+A_2+\tilde A_1+\tilde A_2)+A_3+\tilde A_3;
\nonumber\\
&& \label{Z6'masses}\eea
with $\a=\sqrt{25-128\sqrt{3}{\cal V}_3+768{\cal V}_3^2}$ and
$\b=\sqrt{5(135-384\sqrt{3}{\cal V}_3+1280{\cal V}_3^2)}$. Note
that the eigenvalues are invariant under the T-duality symmetry of
the theory ${\cal V}_2\to 1/4{\cal V}_2$. Thus, all $U(1)$s become
massive, including the two anomaly free combinations.

\subsection{Decompactification of the $Z'_6$ orientifold}

The axions that cancel the anomalies, being twisted RR fields, are
localized on the fixed points of the internal dimensions. Since
there are various orbifold sectors $k$, there are also various
axions $\a^i_k$ localized on the fixed points of the internal tori
where the $k$-th orbifold element acts \cite{Klein:1999im}. Thus, in
the $Z'_6$ orientifold, the $\a^i_1,\a^i_5$ axions are living in
the 4D Minkowski space, the $\a^i_2,\a^i_4$ in 4D Minkowski space plus
the second torus $T_2^2$ and the $\a^i_3$ in 4D Minkowski space
plus the third torus $T_3^2$.

The decompactification limit of the first torus (${\cal V}_1\to
\infty$) does not have any special interest since none of the
fields become six-dimensional.

\subsubsection{Decompactification of the second torus (${\cal V}_2\to \infty$)}

If we decompactify the second torus (${\cal V}_2\to \infty$) the
99 states that are coming from the $k=2,4$ sectors and the
$\a^i_2, \a^i_4$ axions become 6 dimensional fields. The gauge group
is enhanced and can be found by the action of $\g_2$, $\g_4$ on the
Chan-Paton factors. The fields of the other
sectors remain four-dimensional and do not contribute to
six-dimensional anomalies. The 'shift' vector will be $2V_9$,
where $V_9$ is given in (\ref{Z6'vectors}). Following the known
procedure we find that the four-dimensional $U(4)\times U(4)\times
U(8)$ gauge group is enhanced in $U(8)\times SO(16)$. The
generators of the $U(4)_1\times U(4)_2$ are enhanced in the
generators of the $U(8)$ as $T_{U(8)}\sim T_{U(4)_1} \oplus
\overline{T}_{U(4)_2}$ and the generators of the $U(8)$ in the
generators of the $SO(16)$.

The rest of the matter fields are combined with some Kaluza-Klein
states, that now become massless, to give the representations of
the greater gauge group. The $(4,4,1)$, $(\bar{4},\bar{4},1)$ are
now contained in the adjoint of the $U(8)$ as the $(1,1,28)$,
$(1,1,\overline{28})$ are contained in the adjoint of the
$SO(16)$. The $(6,1,1)$, $(1,\bar{6},1)$ form the antisymmetric
$(28,1)$. The $(\bar{4},4,1)$ form the $(\overline{28},1)$.
Finally, the $(4,1,8)$, $(1,\bar{4},\bar{8})$, $(4,1,\bar{8})$ and
$(1,\bar{4},8)$ form the bi-fundamental $(8,16)$.
Thus, the effective gauge group is the one that it was taken from
the $Z_3$ six-dimensional orientifold (Table 1).

The spectrum of the $Z_3$ six-dimensional orientifold contains an
anomalous gauge boson (chapter \ref{Z_3}). By the way that the
$U(4)\times U(4) \times U(8)$ gauge group is enhanced in $U(8)\times
SO(16)$, we find that the anomalous gauge boson is $A_1-A_2$ and becomes
massive due to the six-dimensional Green-Schwarz mechanism.
This mass can be evaluated by the six dimensional formulae and it
is given in (\ref{Z_3massOfU(1)}). The $A_1+A_2$ and $A_3$ are
enhanced in the non-Abelian factors and they have no anomalies.

The contribution of the six-dimensional masses to the
four-dimensional ones can be found by taking the ${\cal V}_2\to
\infty$ limit of (\ref{Z'6.99.D4->D6}):
\bea {1\over 2}M^2_{99,ij}=&-&{1\over 4\pi^3}\left(Tr[\g_2\l^9_i]
Tr[\g_2\l^9_j] +Tr[\g_4\l^9_i]Tr[\g_4\l^9_j]\right)~. \eea
which is the same as the formula of the masses in the
six-dimensional $Z_3$ orientifold (\ref{Z_3massOfU(1)}) upon
normalization. The sectors $k=2,4$ of the four-dimensional $Z_6'$
orientifold in this limit are the $k=1,2$ sectors of the
six-dimensional $Z_3$ orientifold. Using (\ref{lambda_Gen}) and
(\ref{Z6'vectors}), we evaluate the mass-matrix of the anomalous
$U(1)$s. The mass-matrix has two zero eigenvalues, with
eigenvectors: $A_3$, $A_1+A_2$ and a massive state with
eigenvalue:
\be -A_1+A_2~, ~~~~~~~m^2= {3\over \p^3} \ee
as it was expected by the way that the initial $U(4)\times U(4)
\times U(8)$ gauge group is enhanced in $U(8)\times SO(16)$. This
six-dimensional contribution affects the four-dimensional mass
(\ref{MZ6'1}).

The results confirm that anomalous gauge bosons in six-dimensions that
become massive through the six-dimensional Green-Schwarz mechanism,
contribute to the four-dimensional mass generation by a normalized
term.

\subsubsection{Decompactification of the third torus (${\cal V}_3\to
               \infty$)}

If we decompactify the third torus (${\cal V}_3\to \infty$), all
the string states from the $k=3$ sector and the $a^i_3$ axions
become six-dimensional. The new gauge group can be found by the
action of the $\g_3$ on the Chan-Paton. The orbifold rotation
$3(v_1,v_2)=(1,-1)/2$ shows that D5-branes survive in this limit.
The 'shift' vector is now $3V_a$ where $V_a$ is given
in(\ref{Z6'vectors}). The four-dimensional $U(4)_\a\times U(4)_\a\times
U(8)_\a$ gauge group (where $\a=5,9$) is enhanced to $U(16)_\a$ that
is the gauge group of the $Z_2$ six-dimensional orientifold.
The generators are $T_{U(16)}\sim T_{U(4)_1} \oplus T_{U(4)_2}
\oplus \overline{T}_{U(8)}$. Therefore, $(1,\bar{4}, \bar{8})_a$,
$(4,1,8)_a$, $(\bar{4},4,1)_a$ are enhanced in the adjoint of the
$U(16)_a$. The $(6,1,1)_a$, $(1,4,\bar{8})_a$,
$(1,1,\overline{28})_a$, $(4,4,1)_a$ form the antisymmetric
$120_a$. The $(\bar{4},1,8)_a$, $(\bar{4},\bar{4},1)_a$,
$(1,1,28)_a$, $(1,\bar{6},1)_a$ are enhanced in the
$\overline{120}_a$.

From the way that the generators are formed we can expect that the
abelian factor of $U(16)_9$, $A\sim A_1+A_2-\sqrt{2} A_3$ where
the coefficients are coming from the normalization of the
generators of different rank. Similarly for the abelian factor of
$U(16)_5$, $\tilde{A}\sim
\tilde{A}_1+\tilde{A}_2-\sqrt{2}\tilde{A}_3$. As we have seen in
section \ref{Z_2}, the new gauge group contains two anomalous
bosons in six dimensions which are linear combinations of the $A$
and $\tilde{A}$. The other mass eigenstates are embedded in
the non-abelian factors.
The masses of the six-dimensional gauge bosons have been found in
(\ref{Z_2massOfU(1)}). The contribution of the six-dimensional
mass-terms to the four-dimensional mass generation can be found by
taking the ${\cal V}_3\to \infty$ limit in (\ref{Z'6.99.D4->D6}),
(\ref{Z'6.59.D4->D6}) and these are ($a=5,9$):
\bea {1\over 2}M^2_{aa,ij}=&-&{1\over 3\pi^3}Tr[\g_3\l^a_i]
Tr[\g_3\l^a_j]. \label{Mass.Z6'.aa}\eea
and for 59 states:
\bea {1\over 2}M^2_{59,ij}=&-&{1\over 12\pi^3}Tr[\g_3\l^5_i]
Tr[\g_3\l^9_j]. \label{Mass.Z6'.59}\eea
which are the same  (upon normalization) with the contributions of
the six-dimensional generation of the $Z_2$ orientifold (section
\ref{Z_2}). In this limit, the $k=3$ sector of the six-dimensional
$Z_6'$ orientifold is the $k=1$ sector of the six-dimensional
$Z_2$ one. The mass-matrix has four zero eigenvalues, with
eigenvectors: $\sqrt{2} \tilde{A}_1+\tilde{A}_3$,
$-\tilde{A}_1+\tilde{A}_2$, $\sqrt{2} A_1+A_3$, $-A_1+A_2$ and two
massive states with eigenvalues:
\bea A_1+A_2-\sqrt{2}A_3
-\tilde{A}_1-\tilde{A}_2+\sqrt{2}\tilde{A}_3 ~&,&~~~~~~~m_3^2=
{4\over \p^3}
\nonumber\\
-A_1-A_2+\sqrt{2}A_3 -\tilde{A}_1-\tilde{A}_2 +\sqrt{2}\tilde{A}_3
~&,&~~~~~~~m_5^2= {20\over 3\p^3}. \eea
The two massive states are the anomalous $U(1)$ which have been
found in the spectrum of the original six-dimensional $Z_2$
orientifold. The indices are taken from the four-dimensional
counting and denote which masses are affected by six-dimensional
anomalies. Notice that the linear combinations agree with our
expectations.

Another interesting limit of the $Z'_6$ orientifold is ${\cal
V}_3\to 0$. In this limit, the two linear combinations that are
free of four-dimensional anomalies become massless. This is
consistent with the fact that the six-dimensional anomalies which
are responsible for their masses cancel locally in this limit.

\subsection{The four-dimensional $Z_6$ orientifold}

The orbifold rotation vector is $(v_1,v_2,v_3)=(1,1,-2)/6$. Since
there is an order two twist ($k=3$), we have one set of
$D5$-branes that are stretched in the 4D Minkowski and wrap the
third torus $T^2_3$. Tadpole cancellation implies the existence of
32 $D9$-branes and 32 $D5$-branes that we put together at one of
the fixed points of the $Z_2$ action (namely the origin). The
Chan-Paton 'shift' vectors are
\be V_{5,9}={1\over 12}(1,1,1,1,1,1,5,5,5,5,5,5,3,3,3,3)~.
\label{Z6vectors}\ee
The gauge group has a factor of $U(6)\times U(6)\times U(4)$
coming from the $D9$-branes and an isomorphic factor coming from
the $D5$-branes. This orientifold has different supersymmetries in
different sectors. The $N=1$ sectors correspond to $k=1,2,4,5$,
while for $k=3$ we have $N=2$ sectors.

The four-dimensional anomalies of the $U(1)$s have been computed
in \cite{Ibanez:1998qp} and the anomaly matrix is
\be A_{QTT} \sim \left( \ba {cccccc}
 6 & -3 &  \sqrt{6} &  3 &  0 &  \sqrt{6} \\
 3 & -6 & -\sqrt{6} &  0 & -3 & -\sqrt{6} \\
-9 &  9 &      0    & -3 &  3 &     0     \\
 3 &  0 &  \sqrt{6} &  6 & -3 &  \sqrt{6} \\
 0 & -3 & -\sqrt{6} &  3 & -6 & -\sqrt{6} \\
-3 &  3 &      0    & -9 &  9 &     0     \ea \right) \ee
there are three linear combinations that are free of anomalies:
$A_1+A_2-\sqrt{3\over 2}A_3$,
$\tilde{A}_1+\tilde{A}_2-\sqrt{3\over 2}\tilde{A}_3$ and
$A_3-\tilde{A}_3$.

The contributions to the mass matrix \cite{Antoniadis:2002cs} are:
\bea {1\over 2}M^2_{aa,ij}&=&-{\sqrt{3}\over 48\pi^3}
\bigg(Tr[\g_1\l^a_i] Tr[\g_1\l^a_j]
+Tr[\g_5\l^a_i]Tr[\g_5\l^a_j]\nonumber\\
&& +3(Tr[\g_2\l^a_i] Tr[\g_2\l^a_j] +Tr[\g_4\l^a_i]Tr[\g_4\l^a_j])
\bigg)\nonumber\\
&&- {{\cal V}_3\over 3\p^3} Tr[\g_3\l^a_i] Tr[\g_3\l^a_j]
\label{Z6.aa.D4->D6}\eea
for $a=5,9$, while
\bea {1\over 2}M^2_{59,ij}&=&-{\sqrt{3}\over 48\pi^3}
\bigg(Tr[\g_1\l^5_i] Tr[\g_1\l^9_j]
+Tr[\g_5\l^5_i]Tr[\g_5\l^9_j]\nonumber\\
&& +3(Tr[\g_2\l^5_i] Tr[\g_2\l^9_j] +Tr[\g_4\l^5_i]Tr[\g_4\l^9_j])
\bigg)\nonumber\\
&&- {{\cal V}_3\over 12\p^3} Tr[\g_3\l^5_i] Tr[\g_3\l^9_j]
\label{Z6.59.D4->D6}\eea
Notice that the $N=2$ sector contributes with a term proportional
to ${\cal V}_3$.
The mass matrix of the anomalous $U(1)$s has the following
eigenvalues and eigenstates \cite{Antoniadis:2002cs}:
\bea m^2_1=0 &~~~,~~~ & A_1+A_2-\tilde A_1-\tilde
A_2+\sqrt{6}(A_3-\tilde A_3); \\
m^2_2={3\sqrt{3}\over 2} & ~~~,~~~ & A_1-A_2-\tilde A_1+\tilde
A_2; \\
m^2_3=3\sqrt{3} & ~~~,~~~ & A_1-A_2+\tilde A_1-\tilde
A_2; \\
m^2_4=8{\cal V}_3 & ~~~,~~~ & -\sqrt{3\over 2}( A_1+A_2-\tilde
A_1-\tilde A_2)-A_3+\tilde A_3; \\
m^2_{5,6}={7\sqrt{3}+80{\cal V}_3 \pm \b \over 12} & ~~~,~~~ &
{40{\cal V}_3-\sqrt{3}\pm \b \over 12\sqrt{2}-40\sqrt{6}{\cal
V}_3}(A_1+A_2+\tilde
A_1+\tilde A_2)+A_3+\tilde A_3; \nonumber\\
&& \label{Z6masses}\eea
where $\b=\sqrt{147 -1040\sqrt{3}{\cal V}_3+6400{\cal V}_3^2}$. In
the limit ${\cal V}_3\to 0$ the $m_4$, $m_6$ become zero. This is
the consequence of the local cancellation of the six-dimensional
anomalies in this limit.

\subsection{Decompactification of the $Z_6$ orientifold}

In the $Z_6$ orientifold, the $\a^i_1,a^i_2,a^i_4,\a^i_5$ axions
are living in the 4D Minkowski space, and the $\a^i_3$ in 4D
Minkowski space plus the third torus $T_3$.

The decompactification limits of the first and second tori (${\cal
V}_1, {\cal V}_2\to \infty$) do not have any special interest
since none of the fields become six-dimensional and there are no
six-dimensional anomalies.

\subsubsection{Decompactification of the third torus (${\cal V}_3\to
               \infty$)}

If we decompactify the third torus (${\cal V}_3\to \infty$), all
the string states from the $k=3$ sector and the $a^i_3$ axions
become six-dimensional. The rest of the sectors and axions remain
four-dimensional and do not contribute to six-dimensional
anomalies. The new gauge group can be found by the action of the
$\g_3$ on the Chan-Paton. The orbifold rotation
$3(v_1,v_2)=(1,-1)/2$ shows that D5-branes survive in this limit.
The 'shift' vector is now $3V_a$ where $V_a$ is given
in(\ref{Z6vectors}). The old $U(6)\times U(6)\times U(4)$ gauge
group is enhanced to $U(16)$, which is the gauge group of the
$Z_2$ six-dimensional orientifold (Table 1). The generators are
combined as $T_{U(16)}\sim T_{U(6)_1} \oplus T_{U(6)_2} \oplus
\overline{T}_{U(4)}$. Therefore, we can determine how the old
spectrum is enhanced to the new one. The $(\bar{6},1,\bar{4})$,
$(1,6,4)$ and $(6,\bar{6},1)$ combine in the adjoint of
$U(16)$. The $(15,1,1)$, $(1,6,\bar{4})$ are in the antisymmetric
$120$ and $(1,\overline{15},1)$, $(\bar{6},1,4)$ in the
$\overline{120}$.

By the way that the generators of the $U(6)^2\times U(4)$ are
enhanced to the $U(16)$ we can expect that the six-dimensional
$U(1)$ gauge boson of the $U(16)$ will be a linear combination
$A_1+A_2-\sqrt{2 \over 3}A_3$ where the normalization coefficient
in front of $A_3$ takes into account the difference of the rank.
Similarly for the tilde.

The contributions of the six-dimensional anomalies to the
four-dimensional mass generation are given by the ${\cal V}_3\to
\infty$ limit in (\ref{Z6.aa.D4->D6}), (\ref{Z6.59.D4->D6}). We
find (for $a=5,9$):
\be {1\over 2}M^2_{aa,ij}=- {1\over 3\p^3} Tr[\g_3\l^a_i]
Tr[\g_3\l^a_j]\ee
while, for twisted open strings:
\be {1\over 2}M^2_{59,ij}=- {1\over 12\p^3} Tr[\g_3\l^5_i]
Tr[\g_3\l^9_j]\ee
which are the same (upon normalization) as the contributions of
the six-dimensional generation of the $Z_2$ orientifold (section
\ref{Z_2}).
The mass-matrix has four zero eigenvalues, with eigenvectors:
$\sqrt{2\over 3} \tilde{A}_1+\tilde{A}_3$,
$-\tilde{A}_1+\tilde{A}_2$, $\sqrt{2\over 3} A_1+A_3$, $-A_1+A_2$
and two massive states with eigenvalue:
\bea A_1+A_2-\sqrt{2\over 3}A_3
-\tilde{A}_1-\tilde{A}_2+\sqrt{2\over 3}\tilde{A}_3
~&,&~~~~~~~m_4^2= {4\over \p^3}
\nonumber\\
A_1+A_2-\sqrt{2\over 3}A_3 +\tilde{A}_1+\tilde{A}_2-\sqrt{2\over
3}\tilde{A}_3 ~&,&~~~~~~~m_5^2= {20\over 3\p^3}. \eea
The two massive states are the anomalous $U(1)$s which have been
found in the spectrum of the original six-dimensional $Z_2$
orientifold. It is easy to verify that the four-dimensional
massless state $A_1+A_2-\tilde A_1-\tilde A_2+\sqrt{6}(A_3-\tilde
A_3)$ (\ref{Z6masses}) is still massless in six dimensions.

\section{Conclusions}

In this paper we have shown that four-dimensional non-anomalous
$U(1)$s can become massive if in decompactification limits they
suffer from six-dimensional anomalies.

We have studied several four-dimensional orientifolds. In the
decompactification limit, there are sectors in such orientifolds
that become six dimensional.
The original four-dimensional massless spectrum, combined with
Kaluza-Klein states that become massless in this limit, enhanced
to the massless spectrum of six-dimensional orientifolds.
Some RR axions also become six-dimensional fields.

In the six-dimensional orientifolds, we have calculated the
stringy anomalous $U(1)$ masses that are in accordance with
six-dimensional anomalies. The six-dimensional RR axions
contribute to the mass-generation of the anomalous $U(1)$s through
the Green-Schwarz mechanism.

We verified that the six-dimensional mass-matrix is the same as
the volume dependant contribution to the four-dimensional matrix.
Thus, six-dimensional anomalies play indirectly a role in
four-dimensional masses and explain why some non-anomalous $U(1)$
gauge bosons have a non-zero mass.

Our analysis has direct implications for model building both in
string theory and field theory orbifolds. It provides a necessary
and sufficient condition for a non-anomalous $U(1)$ to remain
massless (the hypercharge for example). One has just to check the
associated higher dimensional anomalies in the various
decompactification limits.

The masses of the anomalous $U(1)$s are always as heavy or lighter
than the string scale. Therefore, production of these new gauge
bosons in particle accelerators provides both constrains on model
building and new potential signals at colliders, if the string
scale is around a few TeV.

\newpage

\vskip 1.5cm \centerline{\bf\Large Acknowledgments} \vskip .5cm

The author would like to thank Elias Kiritsis for suggesting the
problem and discussions. He would like also to thank Amine B.
Hammou and Yacine Dolivet for discussions and the Laboratoire de
Physique Th{\'e}orique de l'Ecole Polytechnique for hospitality
during the last stage of this work. The author was partially
suported from RTN contracts HPRN-CT-2000-00122 HPRN-CT-2000-0131
and INTAS contract 99-1-590.

\bigskip\appendix

\section{Definitions and identities}

The Dedekind function is defined by the usual product formula
(with $q=e^{2\pi i\tau}$)
\be \eta(\tau) = q^{1\over 24} \prod_{n=1}^\infty (1-q^n)\ . \ee
The Jacobi $\vartheta$-functions with general characteristic and
arguments are
\be \vartheta [^\a_\b] (z\vert\tau) = \sum_{n\in Z}
e^{i\pi\tau(n-\a/2)^2} e^{2\pi i(z- \b/2)(n-\a/2)}\ . \ee
The $\vartheta[^1_1]$ is an odd function whose first derivative at
zero is $\vartheta[^1_1]^\prime(0|\t) = 2\pi\eta^3$. Some modular
properties of these functions are provided:
\bea \h(-1/\t) = \sqrt{-i\t} \h(\t)\ , \ \ \vartheta
\left[^\a_\b\right] \left({z \over \t} \Bigl| {-1 \over
\t}\right)= \sqrt{-i \t} \ e^{i \p \left({\a \b\over 2} + {z^2
\over \t}\right)} \ \vartheta \left[^{\ \b}_{-\a}\right] (z | \t )
\label{f8} \eea
A very useful identity is
\be \sum_{\alpha,\beta=0,1}\h_{\a\b}~
\vartheta\left[^\a_\b\right](v)\prod_{i=1}^{3}
\vartheta\left[^{\a+h_i}_{\beta+g_i}\right]
(0)=-\vartheta_1(-v/2)\prod_{i=1}^{3} \vartheta
\left[^{1-h_i}_{1-g_i}\right](v/2) \label{SuperID}\ee
valid for $\sum h_i=\sum g_i=0$.

\section{Correlation functions on the annulus}

We present here the derivation of the propagators that we will use
for the calculation of the annulus $\cal A$. This surface can be
defined as quotient of the torus ${\cal T}$ under the involution
\cite{Antoniadis:1996vw}
\be {\cal I}_{\cal A}(z)=1-\bar{z}. \label{involution} \ee

Thus, the correlators can be expressed in terms of the propagators
on the torus. For the bosonic case we have
\be \langle X(z)X(w)\rangle_{\cal T}= -{1 \over 4} \log
\left|\frac{\vartheta_1(z-w|\t)}{\vartheta'_1(0|\t)} \right|^2+ {
\p (z_2-w_2)^2 \over 2 \t_2} \equiv P_B(z,w) \label{bosonicTapp}
\ee
and symmetrizing under the involution:
\bea \langle X(z)X(w)\rangle_{\cal A} &=& {1 \over 2} [ P_B (z,w)+
P_B ({\cal I}_{\cal A} (z),w)+ P_B (z,{\cal I}_{\cal A} (w))+
P_B ({\cal I}_{\cal A} (z),{\cal I}_{\cal A} (w)] \nonumber\\
&=& P_B (z,w)+ P_B (z,1-\bar{w})~. \label{bosonicACapp}\eea
In the amplitude, the partial derivative of the above correlator
(\ref{bosonicACapp}) appears. Thus, we give the expression that we
use for $w=1/2$:
\be \langle \partial_z X(z)X(1/2)\rangle_{\cal A} = -{1 \over 2}
\left[\partial_z\log \vartheta_1(z-1/2|\t) + {2\p i z_2 \over
\t_2}\right]\label{PartialBosonicAC}\ee
for $z=z_1+iz_2$. We remind also that
$\partial_z=(\partial_{z_1}-i\partial_{z_2})/2$.
For the fermionic correlators on the torus we have the identity:
\be \langle\y(z)\y(w)\rangle^2\left[^{\a}_{\b}\right]=-{1\over
4}{\cal P}(z-w)-\p i\partial_{\t} \log{\vartheta
\left[^{\a}_{\b}\right] (0|\t)\over \h(\t)}
\label{FIdentityTapp}\ee
where ${\cal P}(z-w)$ is the Weierstrass function. Symmetrizing
the torus propagator under the involution we find that
(\ref{FIdentityTapp}) holds also for the annulus.

\section{Computations in Type I orientifolds}\label{AnnulusUVchapter}

In the appendix, we give some more details about the computations
of the mass term.

\subsection{Open strings attached on the same kind of branes}

The internal partition function of strings attached on the same
kind of branes is:
\be Z_{int,k}^{aa}[^\a_\b]= \prod^2_{j=1} (-2\sin \p k v_j)
{\vartheta[^{~~~\a}_{\b+2kv_j}](0|\t) \over
\vartheta[^{~~~1}_{1+2kv_j}](0|\t)} ~~~~~~~~\textrm{for a=5,9.}
\label{Zint-A} \ee
After the use of (\ref{SuperID}) and the fact that
$\vartheta[^1_1](0|\t)=0$, we find for the annulus amplitude:
\bea {\cal A}^{aa}_k &=& -{1\over 2N}\int [d\t] \t_2^{1+\d/2}[2\p
\h^3(\t)]^\d \left[ {1 \over 2\p \t^3} 4 \sin^2 {\p k\over
N}\right]\nonumber\\
&=&-{(2\p)^\d \over \p N} \sin^2 {\p k\over N} \int_0^{i \infty}
d\t_2 \t_2^{-2+\d/2} \h^{3\d}(\t_2). \label{Annulus11ab-2-A} \eea
We are interested in the UV limit of the above integral. The
annulus moduli is $\t_2=it/2$:
\bea {\cal A}^{aa,UV}_k &=&-{(2\p)^\d \over \p N} \sin^2 {\p
k\over N} 2^{1-\d/2} \int_0^1 dt ~ t^{-2+\d/2} \h^{3\d}(it/2)
\nonumber\\
&=& -{(2\p)^\d \over \p N} \sin^2 {\p k\over N} 2^{1-\d/2}
\int_0^1 dt ~ t^{-2+\d/2} \left[\left({2\over t}\right)^{1/2}
\h\left({2\over t}\right)\right]^{3\d} \nonumber\\
&=& -{4 \over \p^2 \d N} \left( {8 \over \d}\right)^\d \sin^2 {\p
k\over N} ~ \G(1+\d,\p\d/2). \label{Annulus11ab-UV-A} \eea
where $\G(a,x)$ is the incomplete $\G$-function and $\G(1,0)=1$.

\subsection{Open strings attached on different kind of branes}

Strings attached on different kind of branes have coordinates
$X^a$ with mixed Dirichlet-Neumann boundary conditions. Those
coordinates are half-integer moded and there are no windings or
momenta. The fermionic sectors interchange modes between R and NS
(since the R states should have same modes than the coordinates)
keeping the total fermionic pact unchanged. Thus, the internal
partition function for such strings is:
\be Z_{int,k}^{59}[^\a_\b]= \prod^2_{j=1} {\vartheta
[^{~~\a+1}_{\b+2kv_j}] (0|\t) \over
\vartheta[^{~~~0}_{1+2kv_j}](0|\t)}. \label{Zint-59-A} \ee
Following the same procedure, like in the case of the strings with
the same boundary conditions, we substitute (\ref{Zint-59-A}) in
(\ref{Fab-k}) and after a bit of "thetacology" we find:
\be {\cal A}^{59}_k = -{1\over 2N}\int [d\t] \t_2^{1+\d/2}[2\p
\h^3(\t)]^\d \left[ {1 \over 2\p \t^3}\right]
\label{Annulus11ab-2-A_appendix} \ee
The integral is the same as in the case of the strings having the
same boundary conditions. Using the above result we find:
\bea {\cal A}^{59,UV}_k &=& -{1 \over \p^2 \d N} \left( {8 \over
\d}\right)^\d ~ \G(1+\d,\p\d/2). \label{Annulus11ab-59-UV-A} \eea

\section{The anomaly-free massless spectrum of the N=1
         six-dimensional orientifolds}\label{AnomalyFree}

In the next sections, we will show that the spectrum of the N=1
six-dimensional orientifolds does not suffer from irreducible
non-abelian anomalies .

We reevaluate the spectrum of these models using the 'shift'
vectors that are given for each model. We were especially careful
in distinguishing the representations from the conjugate
representations since it was not clear in the previous literature.
Our results are provided in Table 1. In this section we will prove
that the spectrum does not suffer from irreducible non-abelian
anomalies \cite{Berkooz:1996iz, Scrucca:1999pp}.

Before we continue to the computations, we review the $Z_N$
orientifolds. The Ramond sector $|s_1 s_2 s_3 s_4,ij\rangle
\l_{ji}$ is constrained by GSO projection to have even number of
minus signs between the $s_i$. The orbifold acts on the states
with $a^k_N=e^{{2 \p i k \over N} (J_{67}-J_{89})}$.

For all the orientifolds $Z_N$ with $Z\neq 2$, in the vector
multiplet there are two fermionic fields with $s_1=s_2=\pm 1/2$,
that transform like $(2,1)$ under the space-time
$SO(4)=SU(2)\times SU(2)$. However, the hypermultiplets contain
two spinors $(1,2)$ that one is conjugate of the other under the
gauge group. In the $N=6$ for example, the spinors of 99 and 55
transform like:
\bea s_1=-s_2, ~s_3=-s_4=+1/2   &~~~~~~~~~&
(6,1,1),~(\bar{4},1,8),~(1,4,\bar{8}),~(1,\bar{6},1)
\nonumber\\
s_1=-s_2, ~s_3=-s_4=-1/2   &~~~~~~~~~&
(\bar{6},1,1),~(4,1,\bar{8}),~(1,\bar{4},8),~(1,6,1) \eea
and for 59 string-states we have one fermionic state in $(1,2)$:
\bea s_1=-s_2&~~~~~~~~~&
(4,1,1;\overline{4},1,1)+(\overline{4},1,1;4,1,1)+ \nonumber\\
&& (1,4,1;1,\overline{4},1)+(1,\overline{4},1;1,4,1)+ \nonumber\\
&& (1,1,8;1,1,\overline{8})+(1,1,\overline{8};1,1,8) \eea
Thus, the six-dimensional non-abelian anomalies vanish if we use
the relations between the quartic Casimir of $SU(N)$ in various
representations. We need also \cite{Erler:1993zy}:
\bea Tr[T^4]_{adj}&=& 2 N~tr[T^4]_{\fund}+ 6~tr[T^2]^2_{\fund}
\nonumber\\
Tr[T^4]_{\Yasymm}~&=& (N-8)~tr[T^4]_{\fund} +3~tr[T^2]^2_{\fund}~.
\eea
We will cancel the $Tr[F^4]$, non-abelian anomalies for each
orientifold separately.

\subsection{$Z_2$ Orbifold}

The $Z_2$ orbifold in the type I string theory, gives
$U(16)_9\times U(16)_5$ gauge group. The massless states are given
in Table 1. The spectrum does not suffer from non-abelian
anomalies. The contribution of all the spinors in the $Tr[F^4]$
anomaly of $SU(16)_9$ is:
\bea 2(-2\cdot 16)tr[T^4]+2\cdot 2(16-8)tr[T^4]+(16+16)tr[T^4]=0
\eea
The first term is coming from the $2(1,2)$ spinors of the adjoint.
The second term is coming from the two $(2,1)$ that (only in the
$Z_2$ case) transform similarly, like $120+\overline{120}$. The
last term is coming from the $(1,2)$ state (the 59 string states).
Similarly, we can show that the $Tr[F^4]$ anomaly of the
$SU(16)_5$ vanish too.

\subsection{$Z_3$ Orbifold}

In the type I $Z_3$ orbifold, there are only $D9$-branes,
characterized by the $U(8)\times SO(16)$ gauge group. The
vanishing of the non-abelian anomalies is strait-forward:
\bea  SU(8) &~~:~~& 2(-2\cdot 8)tr[T^4]+
2(8-8)tr[T^4]+(16+16)tr[T^4]=0
\nonumber\\
SO(16) &~~:~~& -2(16-8)tr[T^4]+(8+8)tr[T^4]=0 \eea
The first, second and third term for the $SU(8)$ are the
contribution of the adjoint, antisymmetric and bifundamental of
the spectrum. The first and second terms for the $SO(16)$ are the
adjoint and bifundamentals respectively.

\subsection{$Z_4$ Orbifold}

In the type I $Z_4$ orbifold, there are 32 $D9$ and 32
$D5$-branes, characterized by the $U(8)_9\times U(8)_9\times
U(8)_5\times U(8)_5$ gauge group. The contribution of all the
spinors to the $Tr[F^4]$ anomaly of one of the $SU(8)$ is:
\bea 2(-2\cdot 8)tr[T^4]+2(8-8)tr[T^4]+(8+8+8+8)tr[T^4]=0 \eea
The first, second and third term are the contribution of the
adjoint, antisymmetric and bifundamental of the spectrum. The
coefficient have been explained before.

\subsection{$Z_6$ Orbifold}

In the type I $Z_6$ orbifold, there are 32 $D9$ and 32 $D5$-branes
characterized by the $U(4)\times U(4)\times U(8)$ gauge group for
the $D9$-branes and a isomorphic gauge group for the $D5$-branes.
The contribution of all the spinors in the anomaly of one of the
$SU(4)$ is:
\bea 2(-2\cdot 4)tr[T^4]+2(4-8)tr[T^4]+(8+8+4+4)tr[T^4]=0~. \eea
The first, second and third term are the contribution of the
adjoint, antisymmetric and bifundamental of the spectrum. The
contribution in the anomaly of one of the $SU(8)$ is:
\bea 2(-2\cdot 8)tr[T^4]+(4+4+4+4+8+8)tr[T^4]=0~. \eea
%
%



\begin{thebibliography}{999}






\bibitem{Aldazabal:2000dg}
G.~Aldazabal, S.~Franco, L.~E.~Ibanez, R.~Rabadan and
A.~M.~Uranga,
J.\ Math.\ Phys.\  {\bf 42} (2001) 3103 [arXiv:hep-th/0011073];
%
G.~Aldazabal, S.~Franco, L.~E.~Ibanez, R.~Rabadan and
A.~M.~Uranga,
JHEP {\bf 0102} (2001) 047 [arXiv:hep-ph/0011132];
%
G.~Aldazabal, L.~E.~Ibanez, F.~Quevedo and A.~M.~Uranga,
JHEP {\bf 0008} (2000) 002 [arXiv:hep-th/0005067];
%
D.~Cremades, L.~E.~Ibanez and F.~Marchesano,
arXiv:hep-ph/0212048;
%
D.~Cremades, L.~E.~Ibanez and F.~Marchesano,
Nucl.\ Phys.\ B {\bf 643} (2002) 93 [arXiv:hep-th/0205074];
%
D.~Cremades, L.~E.~Ibanez and F.~Marchesano,
JHEP {\bf 0207} (2002) 022 [arXiv:hep-th/0203160];
%
D.~Cremades, L.~E.~Ibanez and F.~Marchesano,
JHEP {\bf 0207} (2002) 009 [arXiv:hep-th/0201205].

\bibitem{Ibanez:2001nd}
L.~E.~Ibanez, F.~Marchesano and R.~Rabadan,
JHEP {\bf 0111} (2001) 002 [arXiv:hep-th/0105155];





\bibitem{Blumenhagen:2000ea}
R.~Blumenhagen, B.~Kors and D.~Lust,
JHEP {\bf 0102} (2001) 030 [arXiv:hep-th/0012156];
%
R.~Blumenhagen, B.~Kors, D.~Lust and T.~Ott,
Nucl.\ Phys.\ B {\bf 616} (2001) 3 [arXiv:hep-th/0107138];
%
R.~Blumenhagen, B.~Kors, D.~Lust and T.~Ott,
Fortsch.\ Phys.\  {\bf 50} (2002) 843 [arXiv:hep-th/0112015];
%
R.~Blumenhagen, B.~Kors and D.~Lust,
Phys.\ Lett.\ B {\bf 532} (2002) 141 [arXiv:hep-th/0202024];
%
R.~Blumenhagen, V.~Braun, B.~Kors and D.~Lust,
JHEP {\bf 0207} (2002) 026 [arXiv:hep-th/0206038];
%
R.~Blumenhagen, V.~Braun, B.~Kors and D.~Lust,
arXiv:hep-th/0210083.



\bibitem{Cvetic:2002qa}
M.~Cvetic, P.~Langacker and G.~Shiu,
Phys.\ Rev.\ D {\bf 66} (2002) 066004 [arXiv:hep-ph/0205252];
%
M.~Cvetic, P.~Langacker and G.~Shiu,
Nucl.\ Phys.\ B {\bf 642} (2002) 139 [arXiv:hep-th/0206115].



\bibitem{Bailin:2000kd}
D.~Bailin, G.~V.~Kraniotis and A.~Love,
Phys.\ Lett.\ B {\bf 502} (2001) 209 [arXiv:hep-th/0011289];
%
D.~Bailin, G.~V.~Kraniotis and A.~Love,
Phys.\ Lett.\ B {\bf 547} (2002) 43 [arXiv:hep-th/0208103];
%
D.~Bailin, G.~V.~Kraniotis and A.~Love,
Phys.\ Lett.\ B {\bf 553} (2003) 79 [arXiv:hep-th/0210219].



\bibitem{Kokorelis:2002ip}
C.~Kokorelis,
JHEP {\bf 0208} (2002) 018 [arXiv:hep-th/0203187];
%
C.~Kokorelis,
JHEP {\bf 0209} (2002) 029 [arXiv:hep-th/0205147];
%
C.~Kokorelis,
JHEP {\bf 0208} (2002) 036 [arXiv:hep-th/0206108];
%
C.~Kokorelis,
arXiv:hep-th/0210004;
%
C.~Kokorelis,
arXiv:hep-th/0211091.



\bibitem{Antoniadis:2000en}
I.~Antoniadis, E.~Kiritsis and T.~N.~Tomaras,
Phys.\ Lett.\ B {\bf 486} (2000) 186 [arXiv:hep-ph/0004214];
%
I.~Antoniadis, E.~Kiritsis and T.~Tomaras,
Fortsch.\ Phys.\  {\bf 49} (2001) 573 [arXiv:hep-th/0111269].
%
I.~Antoniadis, E.~Kiritsis, J.~Rizos and T.~N.~Tomaras,
Nucl.\ Phys.\ B {\bf 660} (2003) 81 [arXiv:hep-th/0210263].







\bibitem{Kiritsis:2002aj}
E.~Kiritsis and P.~Anastasopoulos,
JHEP {\bf 0205} (2002) 054 [arXiv:hep-ph/0201295].

\bibitem{Ghilencea:2002da}
D.~M.~Ghilencea, L.~E.~Ibanez, N.~Irges and F.~Quevedo,
JHEP {\bf 0208} (2002) 016 [arXiv:hep-ph/0205083].

\bibitem{Ghilencea:2002by}
D.~M.~Ghilencea,
Nucl.\ Phys.\ B {\bf 648} (2003) 215 [arXiv:hep-ph/0208205].




\bibitem{Green:sg}
M.~B.~Green and J.~H.~Schwarz,
Phys.\ Lett.\ B {\bf 149} (1984) 117;
%
M.~B.~Green and J.~H.~Schwarz,
Nucl.\ Phys.\ B {\bf 255} (1985) 93.




\bibitem{Sagnotti:1992qw}
A.~Sagnotti,
Phys.\ Lett.\ B {\bf 294} (1992) 196 [arXiv:hep-th/9210127].


\bibitem{Antoniadis:1996vw}
I.~Antoniadis, C.~Bachas, C.~Fabre, H.~Partouche and T.~R.~Taylor,
Nucl.\ Phys.\ B {\bf 489} (1997) 160 [arXiv:hep-th/9608012].




\bibitem{Klein:1999im}
M.~Klein,
Nucl.\ Phys.\ B {\bf 569} (2000) 362 [arXiv:hep-th/9910143].

\bibitem{Scrucca:2002is}
C.~A.~Scrucca, M.~Serone and M.~Trapletti,
Nucl.\ Phys.\ B {\bf 635} (2002) 33 [arXiv:hep-th/0203190].

\bibitem{Antoniadis:2002cs}
I.~Antoniadis, E.~Kiritsis and J.~Rizos,
Nucl.\ Phys.\ B {\bf 637} (2002) 92 [arXiv:hep-th/0204153].




\bibitem{Bianchi:1990tb}
M.~Bianchi and A.~Sagnotti,
Nucl.\ Phys.\ B {\bf 361} (1991) 519.

\bibitem{Gimon:1996rq}
E.~G.~Gimon and J.~Polchinski,
Phys.\ Rev.\ D {\bf 54} (1996) 1667 [arXiv:hep-th/9601038].

\bibitem{Gimon:1996ay}
E.~G.~Gimon and C.~V.~Johnson,
Nucl.\ Phys.\ B {\bf 477} (1996) 715 [arXiv:hep-th/9604129].


\bibitem{Ibanez:1998qp}
L.~E.~Ibanez, R.~Rabadan and A.~M.~Uranga,
Nucl.\ Phys.\ B {\bf 542} (1999) 112 [arXiv:hep-th/9808139].

\bibitem{Aldazabal:1999nu}
G.~Aldazabal, D.~Badagnani, L.~E.~Ibanez and A.~M.~Uranga,
JHEP {\bf 9906} (1999) 031 [arXiv:hep-th/9904071].




\bibitem{Kiritsis:1997hj}
E.~Kiritsis, ``Introduction to superstring theory,''
arXiv:hep-th/9709062.




\bibitem{Berkooz:1996iz}
M.~Berkooz, R.~G.~Leigh, J.~Polchinski, J.~H.~Schwarz, N.~Seiberg
and E.~Witten,
Nucl.\ Phys.\ B {\bf 475} (1996) 115 [arXiv:hep-th/9605184].


\bibitem{Scrucca:1999pp}
C.~A.~Scrucca and M.~Serone,
Nucl.\ Phys.\ B {\bf 564} (2000) 555 [arXiv:hep-th/9907112].


\bibitem{Erler:1993zy}
J.~Erler,
J.\ Math.\ Phys.\  {\bf 35} (1994) 1819 [arXiv:hep-th/9304104].






\end{thebibliography}
\end{document}